\documentclass[preprint2]{aastex631}

\usepackage{amsmath}

\def\e1i{\epsilon_{1\mathrm{i}}}

\usepackage{blindtext}
\usepackage{hyperref}
\usepackage{amsmath,amssymb}
\usepackage{float}
\usepackage{microtype}
\usepackage{graphicx}
\usepackage{bm}
\usepackage{latexsym}
\usepackage{epsfig}
\usepackage{psfrag}
\usepackage{color}
\usepackage{subfigure}
\usepackage[section]{placeins}
\usepackage{comment}

\def\e1i{\epsilon_{1\mathrm{i}}}

\allowdisplaybreaks[1]

\begin{document}
\thispagestyle{empty}
\hfill\footnotesize KCL-PH-TH/2025-50
\title{An Excited Dark Matter Solution to the MeV Galactic Center Excesses}

\author{Shyam Balaji}
\email{shyam.balaji@kcl.ac.uk}
\affiliation{Physics Department, King’s College London, Strand, London, WC2R 2LS, United Kingdom}
\author{Damon Cleaver}
\email{damon.cleaver@kcl.ac.uk}
\affiliation{Physics Department, King’s College London, Strand, London, WC2R 2LS, United Kingdom}
\author{Pedro De la Torre Luque}\email{pedro.delatorre@uam.es}
\affiliation{Departamento de F\'{i}sica Te\'{o}rica, M-15, Universidad Aut\'{o}noma de Madrid, E-28049 Madrid, Spain}
\affiliation{Instituto de F\'{i}sica Te\'{o}rica UAM-CSIC, Universidad Aut\'{o}noma de Madrid, C/ Nicol\'{a}s Cabrera, 13-15, 28049 Madrid, Spain}

\smallskip
\begin{abstract}

Recent COMPTEL data analysis reveals a $\sim2$ MeV continuum excess whose spatial distribution closely matches the long-standing 511 keV line observed by INTEGRAL/SPI, indicating a common population of low-energy positrons that is difficult to reconcile with known astrophysical sources or standard thermal dark matter (DM). We show that a minimal Excited Dark Matter (XDM) model naturally explains these features. In this scenario a DM particle $\chi$ is inelastically upscattered into an excited state $\chi^*$, followed by de-excitation $\chi^*\to\chi e^+ e^-$ producing $\sim$2 MeV positrons that reproduce the 511 keV line morphology and the COMPTEL MeV continuum. Using a full cosmic-ray (CR) propagation treatment, we obtain an excellent fit for $m_\chi\simeq 1.5$ TeV DM particle with mass-splitting $\Delta m =m_{\chi^*}-m_\chi \simeq 4$ MeV for an inelastic geometric scattering cross section of $\sigma_\textrm{mr}=$ 3--4$\times 10^{-23}\,\textrm{cm}^2$. The same positrons supply a substantial, radially flat contribution to the anomalous Central Molecular Zone (CMZ) ionization rate. This is the first unified treatment of XDM-induced positrons across all three observables, yielding correlated MeV signatures testable by upcoming missions targeting the Galactic MeV band.
\end{abstract}

\section{Introduction}
\label{sec:intro}

The Galactic 511\,keV annihilation line has remained one of the most persistent puzzles in high-energy astrophysics for decades. INTEGRAL/SPI measurements \citep{Prantzos:2010wi,kierans2019positron,Siegert:2019tus,Siegert_2023} reveal a bright, nearly symmetric emission around the Galactic Center (GC) and extending $\sim 6^\circ$–$8^\circ$ on the sky. The inferred positron injection rate, $R_{e^+}\!\sim\!10^{43}\,{\rm s^{-1}}$~\citep{Prantzos:2010wi}, together with the 
 narrow extension of the signal, implies injection energies below a few–tens of MeV.  

While $\beta^+$ decays of $^{26}$Al, $^{44}$Ti and $^{56}$Ni can account for a substantial fraction of the disk component \citep{Diehl:2005py,Prantzos:2010wi}, conventional astrophysical populations struggle to reproduce the bright, compact bulge emission \citep{Siegert_2023}. Stellar templates such as the nuclear stellar bulge or a boxy bulge can be tuned to match the observed morphology~\citep{Siegert:2021trw}, which has been interpreted as support for a stellar origin. However, magnetohydrodynamic simulations of Milky Way–like galaxies show that dark matter (DM)–baryon interactions can drive the DM distribution into a configuration that closely traces the stellar bulge and boxy bulge structure~\citep{Silk}. In such cases, template fits based solely on stellar tracers cannot reliably distinguish a purely stellar explanation from one in which DM contributes significantly.

The DM interpretation of the 511\,keV line has therefore received sustained attention 
\citep{Boehm:2003bt,Vincent:2012an,DelaTorreLuque:2024fcc,DelaTorreLuque:2023cef,laTorreLuquePedro:2024est,Nguyen:2025tkl}. 
Simple $s$-wave annihilation scenarios, in which MeV-scale DM directly produces $e^+e^-$ pairs, are tightly constrained by CMB and Big Bang Nucleosynthesis \citep{Beacom:2005qv,Sizun:2007ds,Wilkinson:2016gsy,DelaTorreLuque:2024fcc}, disfavoring standard thermal MeV DM as the source of the positrons (however, see Ref.~\citep{Aghaie:2025dgl} for an interesting possibility which evades current constraints and is able to reproduce the 511 keV observations).

A particularly promising alternative is the \emph{Excited Dark Matter} (XDM) framework proposed in \citet{Finkbeiner:2007kk}. In this scenario, collisions between DM particles inelastically excite one particle to a slightly heavier state that decays back via a light mediator, producing an $e^+e^-$ pair. Because excitation requires a threshold in relative velocity, the positron production rate acquires an exponential dependence on the local velocity dispersion. As the halo velocity dispersion decreases with galactocentric radius, the upscattering rate is strongly suppressed outside the inner kiloparsec, yielding a bulge-peaked morphology that naturally declines toward the disk. Quantitative studies by \citet{Vincent:2012an} and \citet{Cappiello:2023qwl} showed that such XDM models can reproduce the INTEGRAL/SPI angular profile for TeV-scale DM masses, providing a markedly better fit than standard annihilating DM scenarios.

In parallel, a reanalysis of archival COMPTEL data has uncovered a broad excess around 2–3\,MeV, detected at $\sim 8\sigma$ significance and consistent with in-flight annihilation (IfA) of $\sim 2$~MeV positrons \citep{Knodlseder:2025pnx}. The spatial distribution of this MeV excess closely tracks the 511\,keV emission, strongly suggesting a common source of low-energy positrons. 

A third anomaly arises in the Central Molecular Zone (CMZ), the dense neutral-gas complex within the central few hundred parsecs of the Galaxy. Multiple molecular tracers indicate an H$_2$ ionization rate $\zeta_{\mathrm{H}_2}\gtrsim 10^{-15}\,\mathrm{s^{-1}}$ \citep{Oka_2005,Oka_2020}, which is difficult to reconcile with standard CRs (cosmic rays): they are strongly attenuated in dense clouds and their predicted ionization rate falls short by at least two orders of magnitude \citep{Ravikularaman_2025,Huang_2021}. Moreover, ionization powered by a single central source such as Sgr~A* or HESS J1745-290 is expected to peak sharply near the source and decline steeply with radius, whereas the data favour a much flatter radial profile across the CMZ \citep{Dogiel_2013,Ravikularaman_2025}. \citet{DelaTorreLuque:2024fcc} showed that MeV DM annihilating into $e^\pm$ can efficiently ionize the CMZ and reproduce both the level and approximate flatness of the ionization profile, in a region of parameter space compatible with existing constraints and with positron injection rates comparable to those required for the 511\,keV line.

In this work we show that a minimal XDM model can simultaneously account for all three Galactic observables: the INTEGRAL/SPI 511\,keV morphology, the COMPTEL 2–3\,MeV IfA excess, and a substantial fraction of the anomalous CMZ ionization rate. We focus on TeV-scale $m_\chi$ and few-MeV mass splittings $\Delta m$, where threshold-suppressed inelastic upscattering reproduces the observed bulge-dominated morphology and the associated $\sim 2$~MeV positron population naturally yields both the IfA continuum and an approximately radially flat ionization profile in the CMZ.

We perform, the first full CR propagation analysis of the XDM scenario, implementing the injected $e^\pm$ spectrum in the \textsc{DRAGON2} framework to model transport, cooling and annihilation in the Galactic environment. This framework refines the 511\,keV morphology which we then apply to the higher-energy IfA component, providing the first XDM-based explanation of the $\sim 2$–3\,MeV COMPTEL bump. We then finally compute the contribution of these electrons and positrons to the CMZ ionization rate.

The remainder of this paper is organised as follows. Section~\ref{sec:model} summarizes the XDM mechanism, introduces the key phenomenological parameters and describes our implementation of the excitation rate and positron source term. Section~\ref{sec:result} presents the fit to the 511\,keV line, and the prediction of the IfA signal and the CMZ ionization contribution. Section~\ref{sec:conclusion} discusses the implications of our results and the prospects for testing this scenario with future MeV-band observations.

\section{Excited Dark Matter Framework}
\label{sec:model}

\subsection{Overview and Physical Picture}

We consider a minimal dark sector with two nearly degenerate states: a stable ground state $\chi$ and a slightly heavier excited state $\chi^*$, separated by a small mass splitting $\Delta m = m_{\chi^*} - m_\chi$. This setup, first proposed for the 511~keV line in \citet{Finkbeiner:2007kk} and revisited in \citet{Vincent:2012an,Cappiello:2023qwl}, provides a simple way for DM to produce low-energy positrons in the Milky Way.

Collisions between DM particles can inelastically upscatter one of them to the excited state, $\chi \chi \rightarrow \chi \chi^*$ provided their relative velocity exceeds a kinematic threshold. The excited state then decays $\chi^* \rightarrow \chi e^+ e^-$, releasing an $e^+e^-$ pair with total energy $\sim \Delta m$. In contrast to standard annihilating DM, this mechanism converts a small fraction of the halo \emph{kinetic} energy into electron–positron pairs, leaving the overall DM abundance and large-scale halo structure unchanged. Even if only a tiny fraction of particles are excited per Hubble time, the kinetic energy reservoir is sufficient to power the observed positron production rate in the Galactic bulge \citep{Prantzos:2010wi,Siegert_2023}.

Each upscattering event injects positrons with kinetic energy
$E_{e^+}\simeq \Delta m/2$, so $\Delta m \sim \mathrm{few~MeV}$ naturally yields non-relativistic positrons consistent with the narrow 511~keV line. The strong velocity threshold induces a steep radial dependence in the excitation rate: because the velocity dispersion $\sigma_v(r)$ decreases with galactocentric radius, the upscattering rate is exponentially suppressed beyond the inner kiloparsec. This produces a sharply peaked, bulge-dominated morphology that matches the $\sim 6-8^\circ$ FWHM of the INTEGRAL/SPI signal \citep{Weidenspointner:2007rs,Siegert:2019tus} without requiring ad hoc modifications of the DM density profile.

The excitation process is kinematically inaccessible in the early Universe and does not affect the relic abundance. Freeze-out proceeds through standard annihilation channels, such as $\chi\chi\!\to\!\phi\phi$ or $\chi\chi\!\to\!\mathrm{SM}$, with $\langle\sigma v\rangle_{\rm fo}\simeq 3\times 10^{-26}\,\textrm{cm}^3\textrm{s}^{-1}$, giving the observed DM density. The late-time inelastic channel is then enhanced at halo velocities (for example by Sommerfeld enhancement in the presence of a light mediator) without violating cosmological constraints. To remain in the perturbative regime, the fitted cross section must satisfy the unitarity bound $\sigma_\textrm{mr}\lesssim 4\pi/m_\phi^2$ where $m_\phi$ denotes the light mediator mass.

\subsection{Kinematics and Excitation Threshold}
\label{subsec:kinematics}

The upscattering process $\chi\chi \to \chi\chi^*$ is kinematically allowed only if the relative velocity exceeds a threshold
\begin{equation}
v_{\mathrm{th}} = \sqrt{\frac{4\,\Delta m}{m_\chi}} \, .
\label{eq:vth}
\end{equation}
For benchmark values $m_\chi \sim 1~\mathrm{TeV}$ and $\Delta m \sim 4~\mathrm{MeV}$, one finds $v_{\mathrm{th}}\simeq 1200~\mathrm{km\,s^{-1}}$, which is significantly larger than the velocity dispersion in the inner halo. Only a small fraction of particle pairs in the central few hundred parsecs therefore possess sufficient kinetic energy to excite, and the rate falls steeply with radius as $\sigma_v(r)$ decreases.

This velocity selectivity distinguishes XDM from standard annihilation models. While conventional annihilation scales as $\rho_\chi^2(r)$, the positron production rate acquires an additional exponential cutoff from the velocity distribution,
\begin{equation}
R_{e^+}(r) \propto \rho_\chi^2(r)
\exp\!\left[-\frac{v_{\mathrm{th}}^2}{4\sigma_v^2(r)}\right],
\label{eq:rexp}
\end{equation}
as shown in \citet{Finkbeiner:2007kk,Vincent:2012an,Cappiello:2023qwl}. The exponential factor arises from integrating the inelastic cross section over the local Maxwell–Boltzmann velocity distribution and naturally suppresses the signal outside the Galactic bulge.

At the microscopic level, we parameterize the inelastic upscattering cross section as \citep{Cappiello:2023qwl}
\begin{equation}
\sigma v_{\mathrm{rel}} =
\begin{cases}
\sigma_\textrm{mr}\,\sqrt{v_{\mathrm{rel}}^2 - v_{\mathrm{th}}^2}, & v_{\mathrm{rel}} > v_{\mathrm{th}}, \\[4pt]
0, & v_{\mathrm{rel}} \le v_{\mathrm{th}},
\end{cases}
\label{eq:sigmav}
\end{equation}
where $\sigma_\textrm{mr}$ sets the normalization of the inelastic scattering cross section in the moderately relativistic limit. The threshold $v_{\mathrm{th}}$ introduces a sharp kinematic cutoff, confining excitations to the high-velocity tail of the halo distribution and imprinting the exponential suppression with radius seen in Eq.~(\ref{eq:rexp}). This form approximates the more general expressions obtained in light-mediator models \citep{Finkbeiner:2007kk,Pospelov:2007xh,Arkani-Hamed:2008hhe}, while capturing the astrophysically relevant dependence needed to model the Galactic signal.

In what follows, we fold this threshold behaviour into a full computation of the velocity-averaged excitation rate using the local velocity dispersion $\sigma_v(r)$ derived from the Jeans equation (Section~\ref{sec:morphology}). This quantity controls both the radial morphology of the 511~keV emission and the total positron injection rate relevant for the IfA and CMZ analyses.

\subsection{Positron Injection Rate}
\label{subsec:injection}

The local positron production rate per unit volume from inelastic DM upscattering is
\begin{equation}
Q_{e^+}(r, E_e)
   = \frac{\rho_\chi^2(r)}{2m_\chi^2}\,
     \langle\sigma v\rangle(r) \frac{dN}{dE}(E_e)\, ,
\label{eq:sourceQ}
\end{equation}
where $\rho_\chi(r)$ is the DM density profile, $m_\chi$ is the particle mass, and the spectrum $dN/dE_e = 2\delta(E-\Delta m/2)$ for monoenergetic injection. The factor of $1/2$ avoids double-counting identical initial states. This source term drives all of the subsequent phenomenology.

\vspace{2mm}
\noindent\textbf{Velocity-averaged excitation rate.}  
Following \citet{Cappiello:2023qwl}, the velocity-averaged inelastic excitation cross section is obtained by integrating the microscopic cross section over the local DM velocity distribution,
\begin{widetext}
\label{eq:fullcrosssection}
    \begin{align}
    \langle\sigma v\rangle &= \sigma_\textrm{mr} 
    2\pi\left(\frac{1}{N_{esc}}\frac{1}{(2\pi\sigma_v^2)^{3/2}}\right)^2 \nonumber \\ & \times \int_{v_\textrm{th}}^{2v_\textrm{esc}}\textrm{d}v_\textrm{rel} \ v_\textrm{rel}\sqrt{v_\textrm{rel}^2 - v_\textrm{th}^2}e^{-\frac{2v_\textrm{esc}^2+v_\textrm{rel}^2}{2\sigma_v^2}}\sigma_v^3  
    \left[2\sigma_v(e^{\frac{v_\textrm{rel}^2}{2\sigma_v^2}} - e^{\frac{v_\textrm{esc}v_\textrm{rel}}{2\sigma_v^2}}) + \sqrt{\pi}v_\textrm{rel}e^{\frac{4v_\textrm{esc}^2 + v_\textrm{rel}^2}{4\sigma_v^2}}\textrm{erf}\left(\frac{2v_\textrm{esc}-v_\textrm{rel}}{2\sigma_v}\right)\right]\,, \label{eq:bigsv}
    \end{align}
\end{widetext}
where $\sigma_v(r)$ and $v_{\mathrm{esc}}(r)$ are the local DM velocity dispersion and escape velocity, and $N_{\mathrm{esc}}$ normalizes the truncated Maxwellian distribution. The threshold velocity $v_{\mathrm{th}}$ is defined in Eq.~(\ref{eq:vth}).\footnote{In Eq.~(5) of \citet{Cappiello:2023qwl}, a cross section $\langle\sigma v\rangle_{\rm mr}\simeq \sigma_\textrm{mr} v_{\rm eff}$ was introduced as a constant normalization with an effective halo velocity $v_{\rm eff}\sim 10^{-3} c$. Here we instead derive $\langle\sigma v\rangle(r)$ by explicitly integrating over the full velocity distribution, which is appropriate for a geometric cross section $\sigma_{\rm mr}$ and ensures a self-consistent radial dependence.}


\subsection{Velocity Dispersion and Escape Velocity}
\label{sec:morphology}

\begin{figure}
    \centering
    \includegraphics[width=0.99\linewidth]{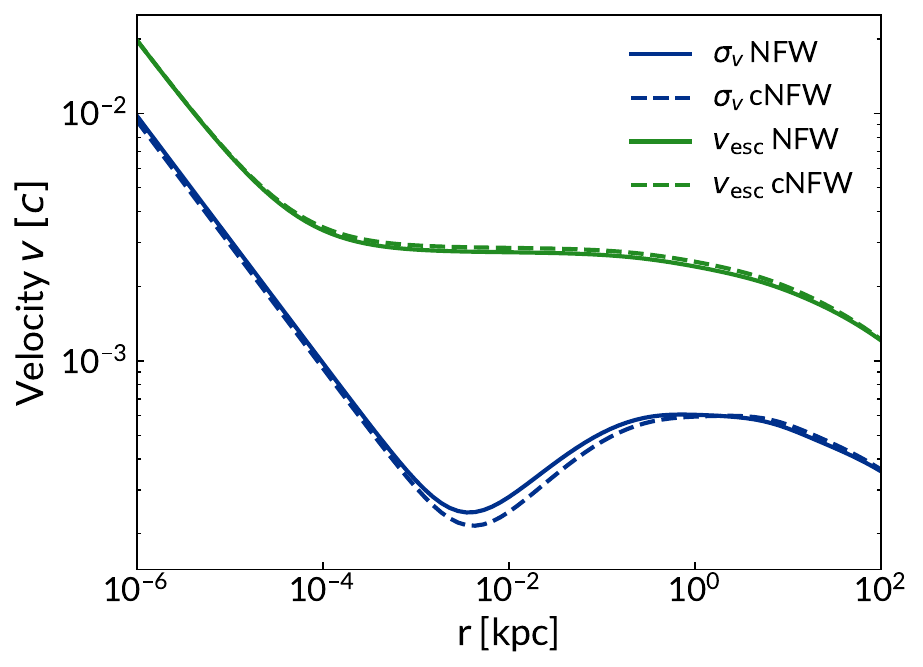}
    \caption{DM velocity profiles as a function of radial distance from the GC for the velocity dispersion $\sigma_v$ (blue) and escape velocity $v_{\mathrm{esc}}$ (green) for NFW (solid) and cNFW (dashed) DM density distributions. The cNFW profile has inner slope $\gamma=1.2$.}
    \label{fig:velocities}
\end{figure}

The velocity dispersion $\sigma_v(r)$ and escape velocity $v_{\mathrm{esc}}(r)$ entering Eq.~(\ref{eq:bigsv}) are shown in Fig.~\ref{fig:velocities}. They are computed by solving the isotropic Jeans equation under spherical symmetry, following \citet{Vincent:2012an,Cappiello:2023qwl}. The total gravitational potential includes contributions from DM, bulge, disk and Sgr~A*. We adopt the same baryonic potential models and parameter choices as \citet{Vincent:2012an}, which reproduce Milky Way rotation-curve data and yield $\sigma_v \sim 200$–$250~\mathrm{km\,s^{-1}}$ and $v_{\mathrm{esc}} \sim 600$–$700~\mathrm{km\,s^{-1}}$ in the inner kiloparsec.

\subsection{511 keV Morphology and Positron Propagation}
\label{subsec:511}

The angular distribution of 511~keV photons is obtained by integrating the local emissivity of diffuse positrons along the line of sight. Unlike previous morphology studies that assumed instantaneous annihilation at the production site~\citep{Vincent:2012an,Cappiello:2023qwl}, we propagate positrons with non-zero kinetic energy using the \textsc{DRAGON2} code, following \citet{DelaTorreLuque:2023cef}. This includes diffusion, reacceleration and energy losses, and yields a steady-state positron distribution before annihilation.

The resulting flux per unit solid angle over line-of-sight distance $s$ for a position $(x,y,z)$ is
\begin{equation}
\frac{d\phi_{\gamma}^{511}}{d\Omega}
= 2\,k_{\mathrm{ps}}\int ds\,s^{2}\,
\frac{\epsilon_{e}(x_{s,b,l},y_{s,b,l},z_{s,b,l})}{4\pi s^{2}},
\label{eq:511flux}
\end{equation}
where $k_{\mathrm{ps}} = 1/4$ is the fraction of positronium that decays into para-positronium producing two 511~keV photons. The quantity $\epsilon_{e}$ is the line emissivity,
\begin{equation}
\epsilon_{e} = \frac{d\phi_{e}}{dE}(E_{\mathrm{th}})\,
n_{e}\,
\sigma^{\mathrm{ps}}(E_{\mathrm{th}}),
\end{equation}
where $\frac{{d\phi_{e}}}{{dE}}(E_{\mathrm{th}})$ is the propagated positron flux at thermal energy $E_{\mathrm{th}}$, $\sigma^{\mathrm{ps}}$ is the charge-exchange cross section for positronium formation with hydrogen~\citep{JeanP_2009}, and $n_{e}$ is the ambient electron density. We take $n_{e}=1~\mathrm{cm^{-3}}$ in the Galactic plane and adopt an exponential decrease above and below the plane as in \citet{laTorreLuquePedro:2024est}, combining a Nakanishi warm-gas component~\citep{Nakanishi_2003} and a Ferriere hot-gas component~\citep{Ferriere_1998}.

This treatment correctly accounts for the finite propagation length of injected MeV positrons, producing a slightly more extended bulge emission than instantaneous-annihilation models. 

\subsection{In-Flight Annihilation Excess}
\label{subsec:IFA}

Relativistic positrons injected in the interstellar medium can annihilate with electrons before becoming thermal, producing a continuum of $\gamma$ rays up to the positron energy. IfA has long been recognized as a probe of MeV DM scenarios \citep{Stecker,Beacom:2005qv,Sizun:2006uh}, since high injection energies generate an IfA flux that can overshoot the observed diffuse MeV background \citep{COMPTEL1994,Bouchet_2011}. 


As discussed in Sec.~\ref{sec:intro}, the reanalysis of COMPTEL data by \citet{Knodlseder:2025pnx} reveals a MeV excess consistent with IfA of $\sim 2$ MeV positrons sharing the 511 keV spatial template. In our XDM scenario, these positrons are produced in $\chi^* \to \chi e^+e^-$ decays with injection energy $E_{e^+} \simeq \Delta m/2$. For monoenergetic injection, the differential IfA $\gamma$-ray flux can be related to the 511~keV intensity \citep{Beacom:2005qv,Beacom:2004pe}:
\begin{equation}
\frac{d\Phi^{\mathrm{IfA}}}{dE_\gamma}
= \frac{\Phi^{511}}{1 - f_{\mathrm{Ps}}/4}\,
P^{\mathrm{IfA}}(E_\gamma;E_0,n_e),
\label{eq:IFA_flux}
\end{equation}
where $f_{\mathrm{Ps}}\simeq 0.97$ is the positronium fraction~\citep{JeanP_2009,Bouchet:2010dj} and $P^{\mathrm{IfA}}(E_\gamma;E_0,n_e)$ is the photon spectrum per injected positron of initial energy $E_0 = \Delta m/2$ in a medium of density $n_e$.

Here we compute the IfA signal following \citet{laTorreLuquePedro:2024est}, using the steady-state positron spectrum obtained from \textsc{DRAGON2}:
\begin{equation}
\begin{aligned}
\frac{d\Phi^\textrm{IfA}}{d\Omega\, dE_{\gamma}}
&= \frac{d\Phi^{511}}{d\Omega}
   \frac{n_H}{P\!\left(1-\frac{3}{4}f_{\mathrm{Ps}}\right)}
\\
&\quad \times
\int_{E_{\gamma}=m_e/2}^{E_\textrm{max}} 
    dE'\,\frac{1}{N_{\rm pos}}
    \frac{dN_{\rm pos}}{dE'}
\\
&\quad \times
\int_{m_e}^{E'} 
    P_{E'\rightarrow E}\,
    \frac{d\sigma}{dE_{\gamma}}
    \frac{dE}{|dE/dx|}\,,
\end{aligned}
\label{eq:IfA}
\end{equation}
where the first integral over $E'$ accounts for the propagated positron energy distribution, bounded between $E_\textrm{max}$ (the maximum steady-state energy) and $m_e/2$~\citep{Bartels:2017dpb}. The factor $(1/N_{\rm pos})\,dN_{\rm pos}/dE'$ gives the fraction of positrons at energy $E'$ that contribute to the 511~keV emission. The hydrogen density $n_\textrm{H}$ sets the number of target electrons, $P_{E'\to E}$ is the probability for a positron to produce a photon before cooling from $E'$ to $E$ \citep{Beacom:2005qv}, and $P=P_{E'\to m_e}$ is the total probability to emit a photon before reaching thermal energies.

\subsection{Ionization of the Central Molecular Zone}
\label{subsec:CMZ}



As discussed in Sec.~\ref{sec:intro}, multiple molecular tracers indicate an H$_2$ ionization rate in the CMZ of $\zeta_{\mathrm{H}_2}\gtrsim 10^{-15}\,\mathrm{s^{-1}}$ with an approximately flat radial profile \citep{Oka_2005,Oka_2020,Dogiel_2013,Ravikularaman_2025}, which is difficult to explain with standard CRs alone \citep{Huang_2021,Ravikularaman_2025}. In light of the MeV–DM explanation proposed by \citet{DelaTorreLuque:2024fcc}, we now ask whether the same XDM setup that fits the 511 keV morphology and IfA signal can also contribute appreciably to the CMZ ionisation.

We follow \citet{DelaTorreLuque:2024fcc} to compute the H$_2$ ionization rate $\zeta_{\textrm{H}_2}$ caused by the $e^\pm$ pairs emitted in XDM de-excitations inside the CMZ. The rate is
\begin{equation}
  \zeta_{\textrm{H}_2}= 2\cdot 4\pi \int^{E_\textrm{max}}_{E_\textrm{min}} J(E, \textbf{x})\sigma(E)(1 + \theta_e(E)) dE,
\label{eq:main}
\end{equation}
where the prefactor 2 accounts for ionization by both $e^-$ and $e^+$, $E_\textrm{max} = \Delta m/2$ is the injection energy, and $E_\textrm{min}$ is the threshold for ionization in H$_2$ ($15.43$~eV~\citep{Padovani_2009}). The cross section $\sigma(E)$ is the electron–H$_2$ ionization cross section, calculated as in Eq.~(7) of \citet{Padovani_2009}, and $\theta_e(E)$ is the number of secondary ionizations per primary ionization, from Eq.~(2.23) of \citet{krause2015crimecosmicray}. The flux $J(E,\textbf{x})$ is the propagated $e^\pm$ flux from \textsc{DRAGON2}, which includes the strong energy losses suffered by these particles in the CMZ. Typical propagation distances before thermalization are of order tens of parsecs, depending on the injection energy.

This formalism allows us to compute the radial profile of the XDM-induced ionization rate across the CMZ and to quantify its contribution relative to the observed anomaly.

\section{Results}
\label{sec:result}

\subsection{Fit to the 511 keV Morphology}

\begin{figure*} 
    \centering
    \includegraphics[width=0.49\linewidth]{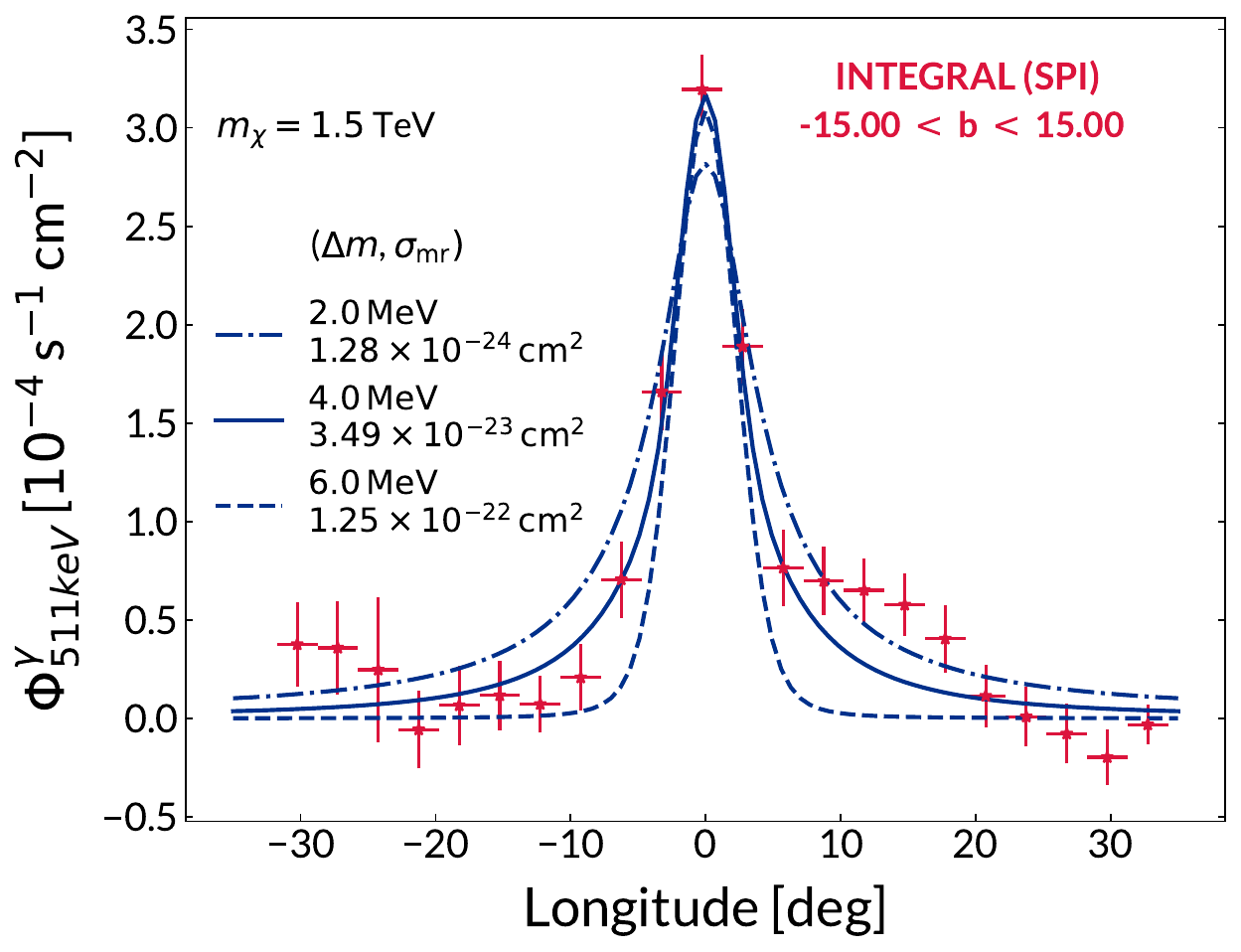}
    \includegraphics[width=0.49\linewidth]{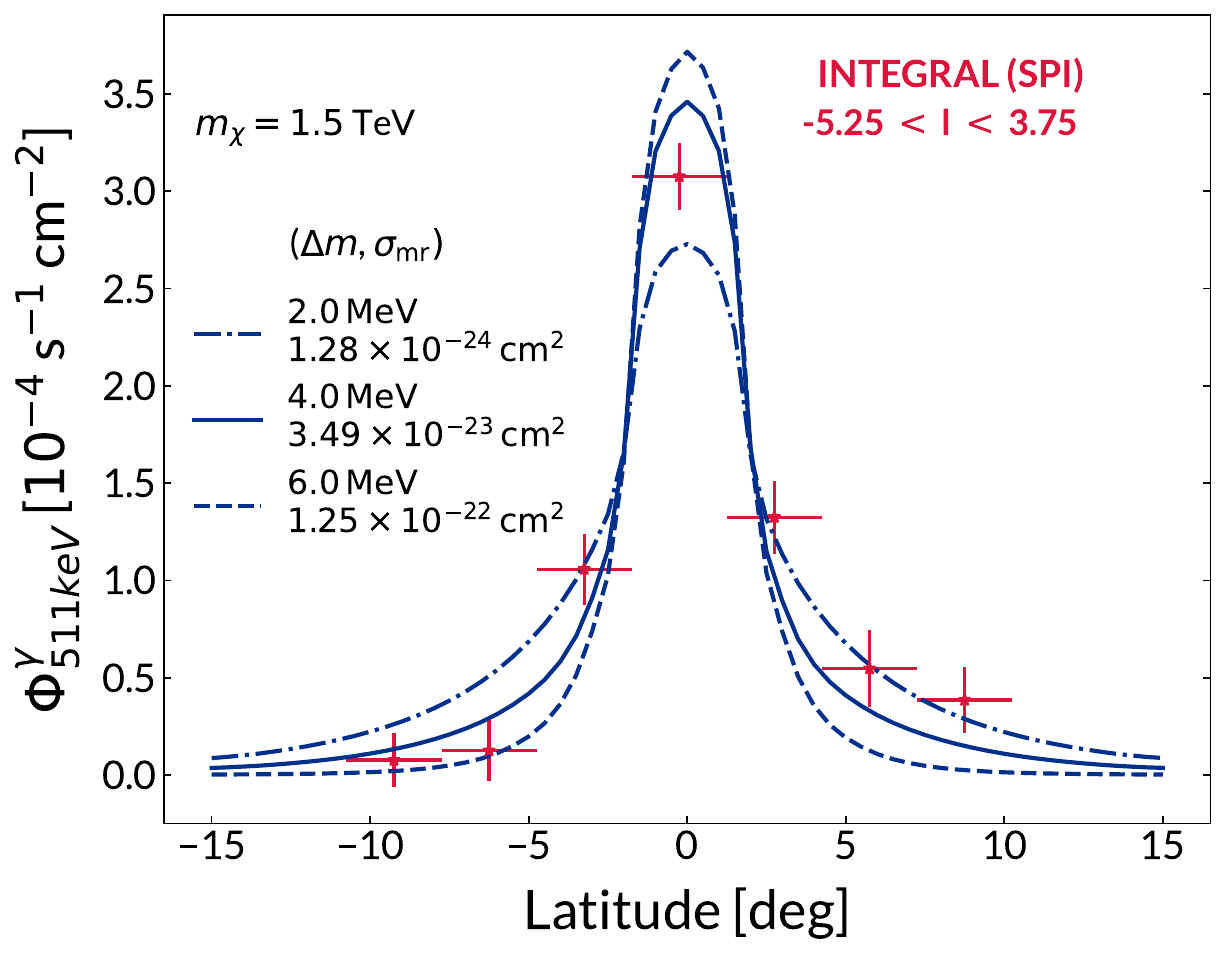}
    \includegraphics[width=0.49\linewidth]{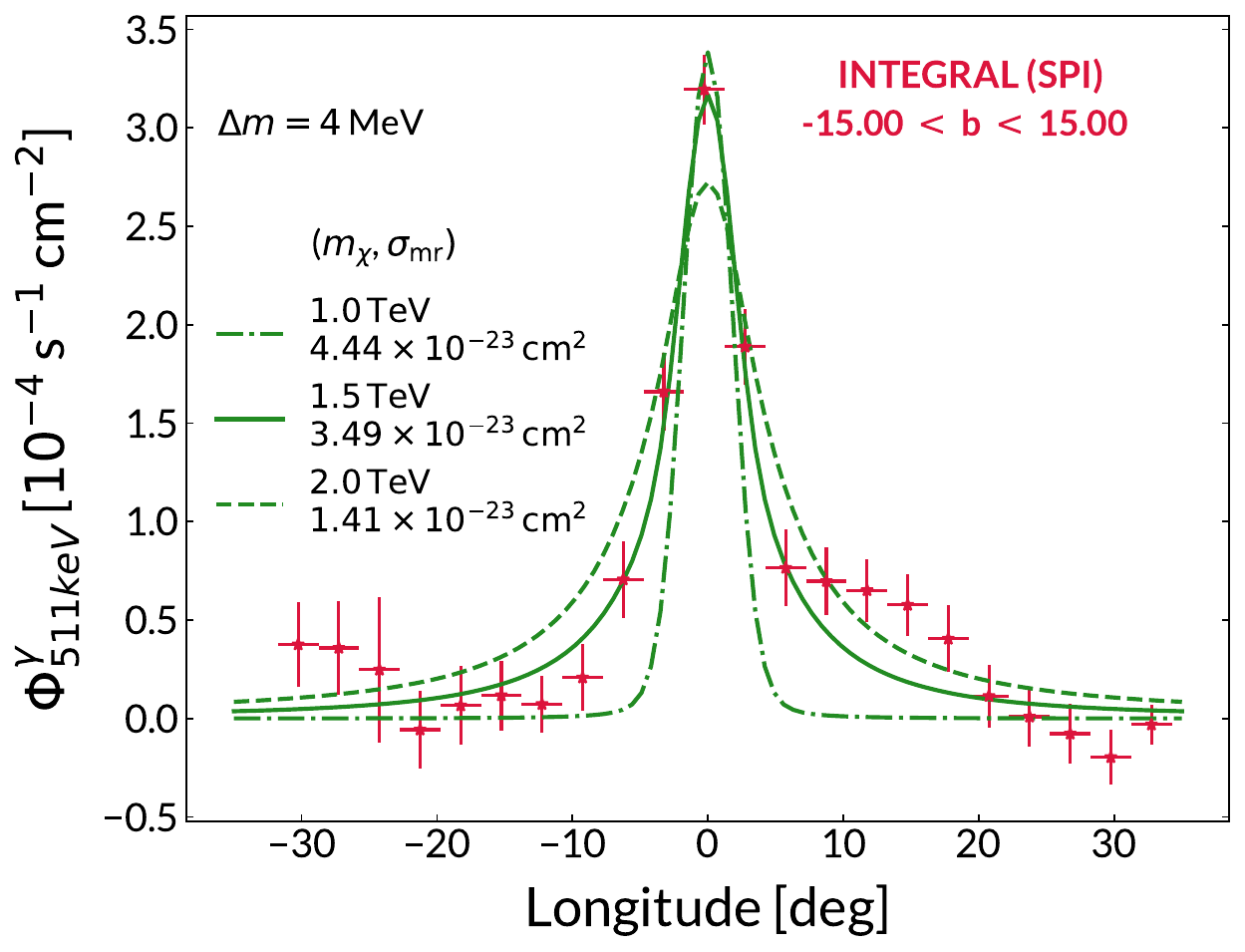}
    \includegraphics[width=0.49\linewidth]{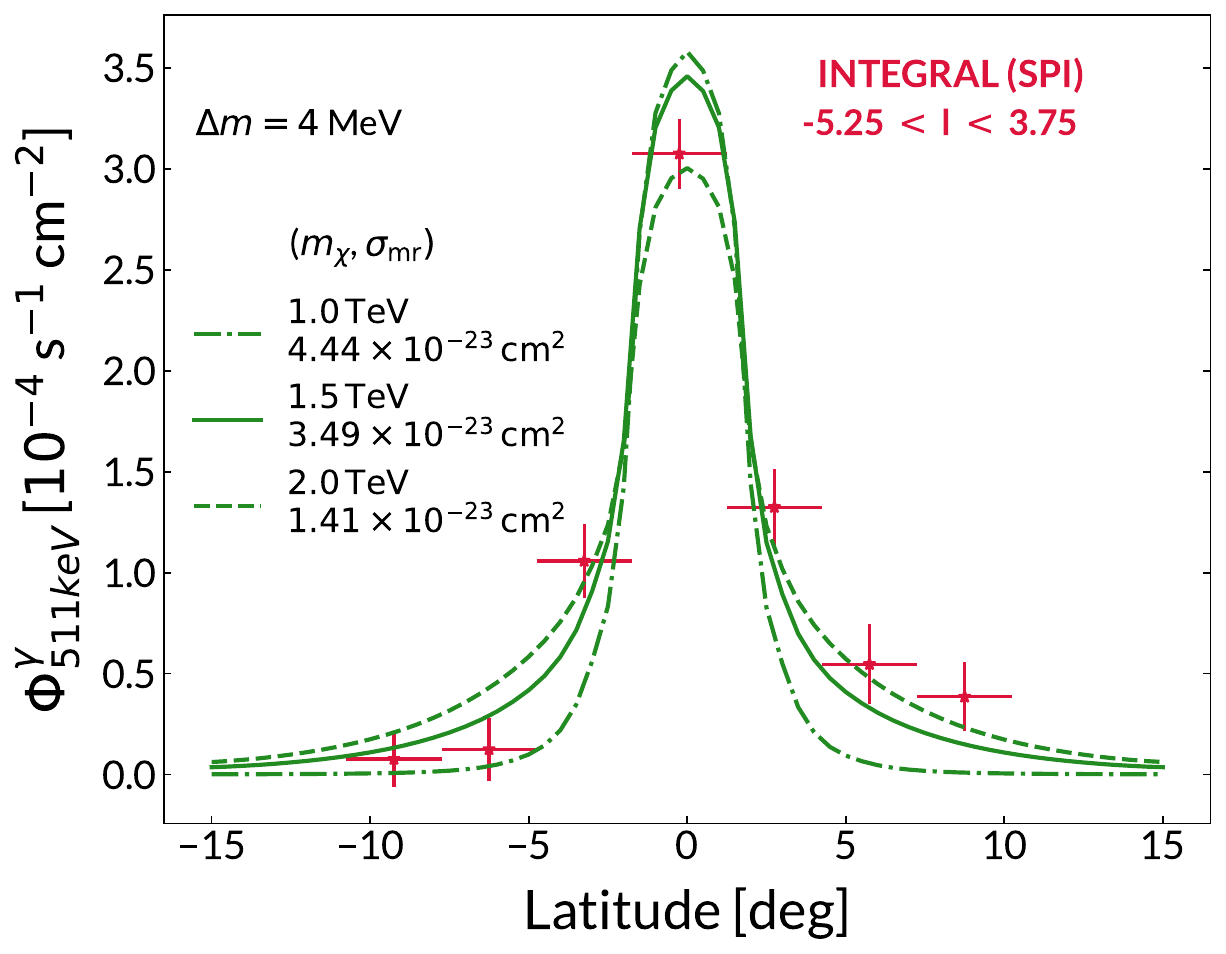}   
    \caption{511 keV longitude profiles (left panels) and latitude profiles (right panels) for XDM models compared to INTEGRAL/SPI data (red points). Top: fixed $m_\chi = 1.5$ TeV and varying mass splittings $\Delta m$. Bottom: fixed $\Delta m = 4$ MeV and varying $m_\chi$.}
    \label{fig:511_best_fit_lat} 
\end{figure*}

We compare the 511 keV flux from Eq.~\eqref{eq:511flux} with INTEGRAL/SPI data, which provide fluxes binned in Galactic longitude $l$ and latitude $b$. Following \citet{Siegert_2023}, we construct a longitude profile by binning in $3^\circ$ intervals in $l$ and integrating over $-10.75^\circ < b < 10.25^\circ$, and a latitude profile by binning in $3^\circ$ intervals in $b$ and integrating over $-5.25^\circ < l < 3.75^\circ$. For a given XDM model, we define
\begin{align}
    \chi^2 &= \sum_b \frac{\big[\phi_{\gamma,b}^{511}(\sigma_{\mathrm{mr}}) - \phi_{\mathrm{SPI},b}\big]^2}{\sigma_{\mathrm{SPI},b}^2} \nonumber\\
    &\quad + \sum_l \frac{\big[\phi_{\gamma,l}^{511}(\sigma_{\mathrm{mr}}) - \phi_{\mathrm{SPI},l}\big]^2}{\sigma_{\mathrm{SPI},l}^2},
\end{align}
where $\phi_{\gamma,i}^{511}(\sigma_{\mathrm{mr}})$ is the predicted XDM flux in bin $i$, and $\phi_{\mathrm{SPI},i}$ and $\sigma_{\mathrm{SPI},i}$ are the measured flux and uncertainty. Varying $\sigma_{\mathrm{mr}}$ we obtain the best-fit value for each choice of $(m_\chi,\Delta m)$, which is then used to compute the IfA spectrum and CMZ ionization.

\begin{figure*}
    \centering
    \includegraphics[width=0.50\linewidth]{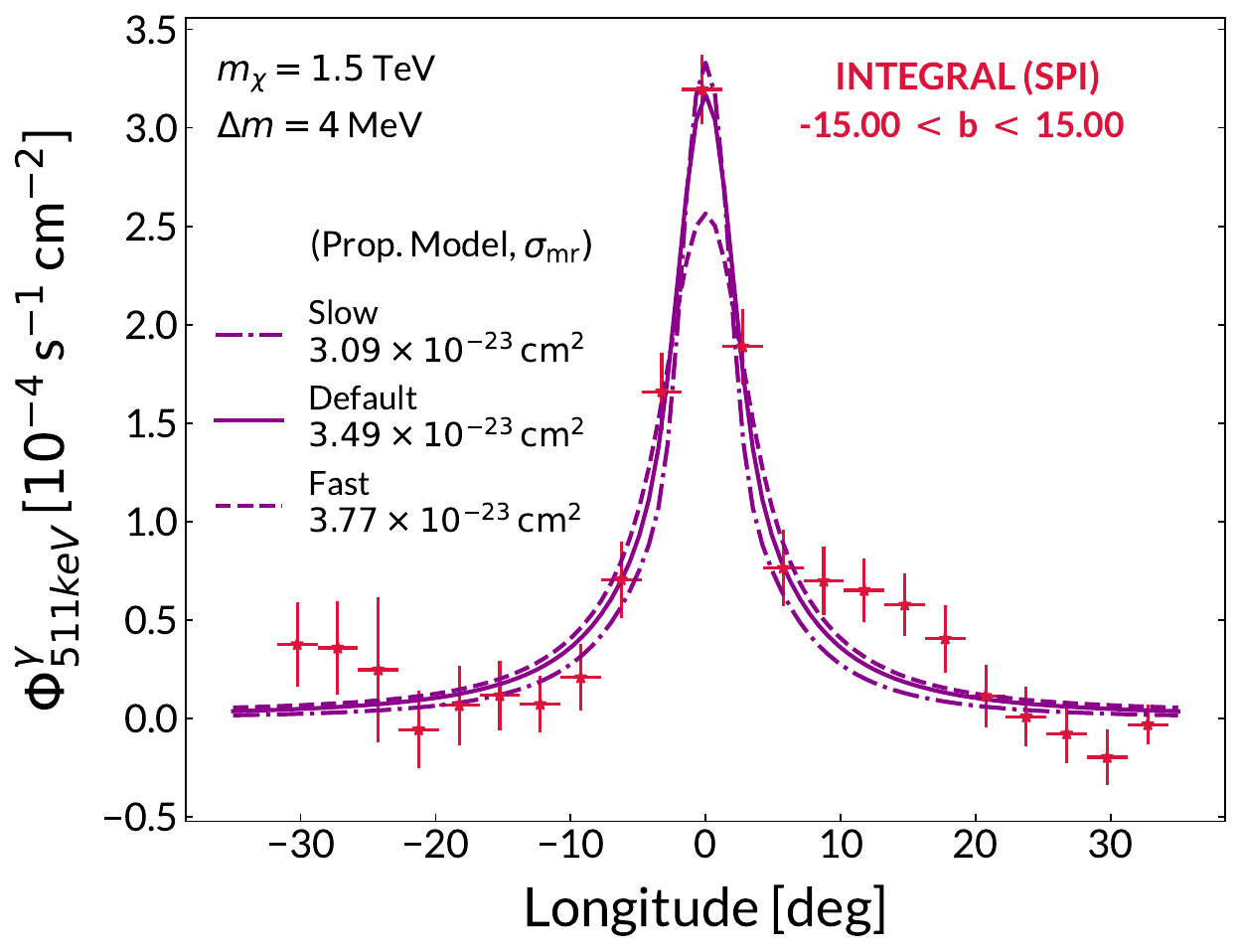}
    \includegraphics[width=0.49\linewidth]{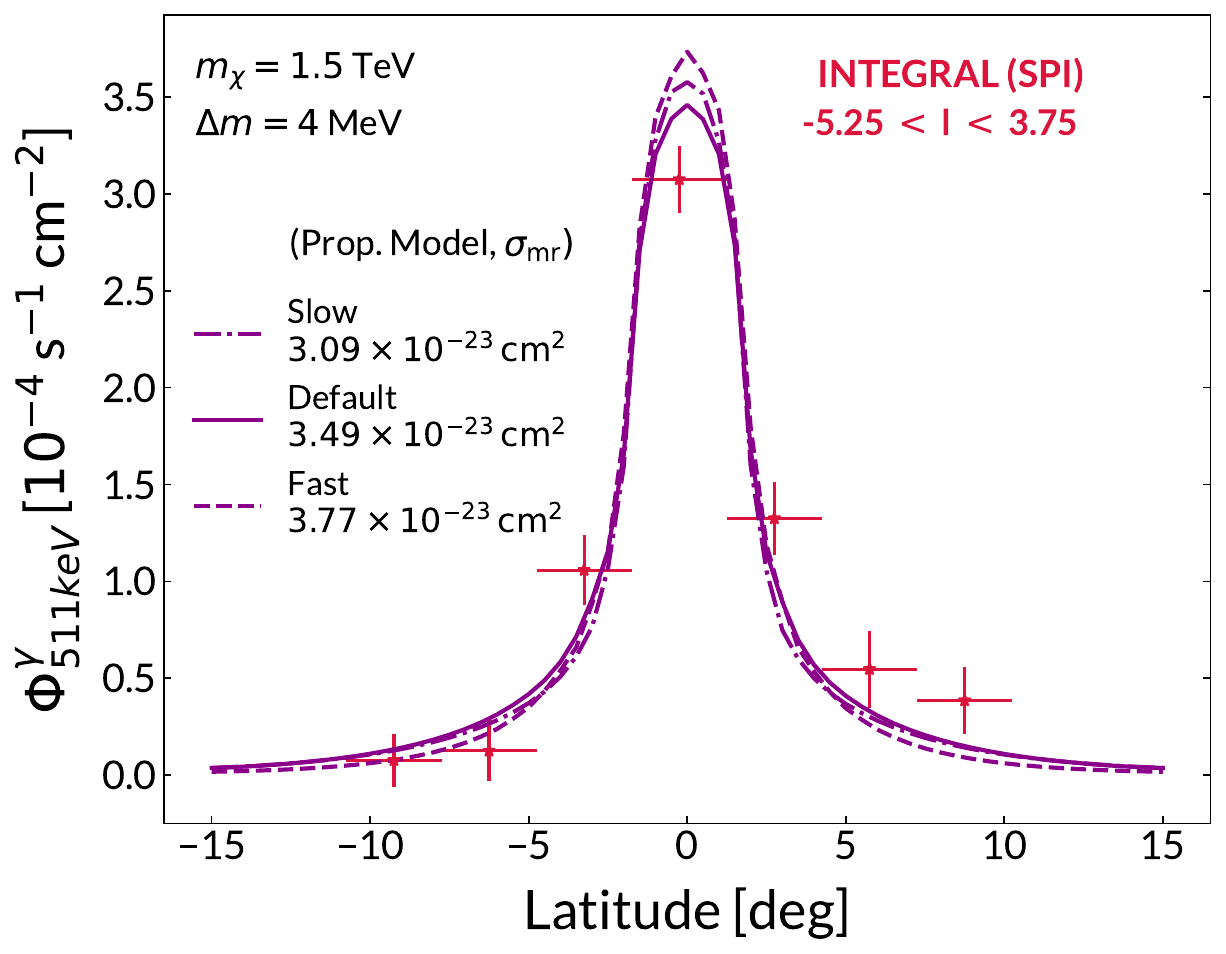}
    \caption{511 keV longitude (left) and latitude (right) profiles for XDM with $m_\chi = 1.5$ TeV and $\Delta m = 4$ MeV, compared to INTEGRAL/SPI data (red points), for three propagation setups: ``Default'' (solid), ``Fast'' (dashed) and ``Slow'' (dash-dotted), following \citet{DelaTorreLuque:2023cef}.}
    \label{fig:511Uncs}
\end{figure*}

The top panels of Fig.~\ref{fig:511_best_fit_lat} show the effect of varying the mass splitting between $2$ and $6$ MeV for fixed $m_\chi = 1.5$ TeV. All of these splittings can, in principle, produce an IfA signal compatible with COMPTEL, but they imprint distinct 511 keV morphologies. The dominant effect is the change in the threshold velocity $v_{\rm th}$, which controls the radial range over which upscattering is efficient. A secondary effect arises from the different positron injection energies, which slightly modify the diffusion length, since the diffusion coefficient is energy dependent.

The bottom panels of Fig.~\ref{fig:511_best_fit_lat} show the longitude and latitude profiles for fixed $\Delta m = 4$ MeV and varying $m_\chi$. This choice of splitting produces monoenergetic positrons with $E_{e^\pm} = \Delta m/2 = 2$ MeV, which is the best-fit injection energy for the bulge component of the IfA spectrum in \citet{Knodlseder:2025pnx}. For $m_\chi = 1.5$ TeV and $\Delta m = 4$ MeV we obtain an excellent fit to the SPI morphology with a reduced chi-squared of $\chi_{\mathrm{red}}^2 \simeq 1.5$ for an NFW profile, even without invoking the IfA or CMZ constraints.

For fixed $\Delta m$ and an NFW density profile, increasing $m_\chi$ broadens the emission. Eq.~\eqref{eq:vth} shows that higher masses lowers the inelastic threshold, so upscattering remains kinematically allowed out to larger radii where the velocity dispersion is smaller. This enhances the velocity-averaged excitation rate coefficient $\langle\sigma v\rangle(r)$ at larger $r$, and smooths the longitude and latitude profiles. Lower masses, by contrast, produce cuspier 511 keV profiles more strongly concentrated toward the GC.

At low $m_\chi$ the predicted bulge emission peaks sharply around the GC and can still be made compatible with SPI data once a disk component from astrophysical positron sources is included, such as massive stars or SNe, following \citet{Cappiello:2023qwl,laTorreLuquePedro:2024est}. Such a disk component is guaranteed at some level, although its modelling carries significant uncertainties. Conversely, a cuspier DM profile naturally yields a cuspier 511 keV morphology and allows larger $m_\chi$ to remain compatible with SPI. This is illustrated in Fig.~\ref{fig:511cNFW}, where we compare the NFW case to a contracted NFW (cNFW) profile with inner slope $\gamma = 1.2$. In this case the density profile, the velocity dispersion and escape velocity (Fig.~\ref{fig:velocities}), and thus the radial dependence of $\langle\sigma v\rangle$ from Eq.~\eqref{eq:bigsv}, are all modified.

\begin{figure*}
    \centering
    \includegraphics[width=0.49\linewidth]{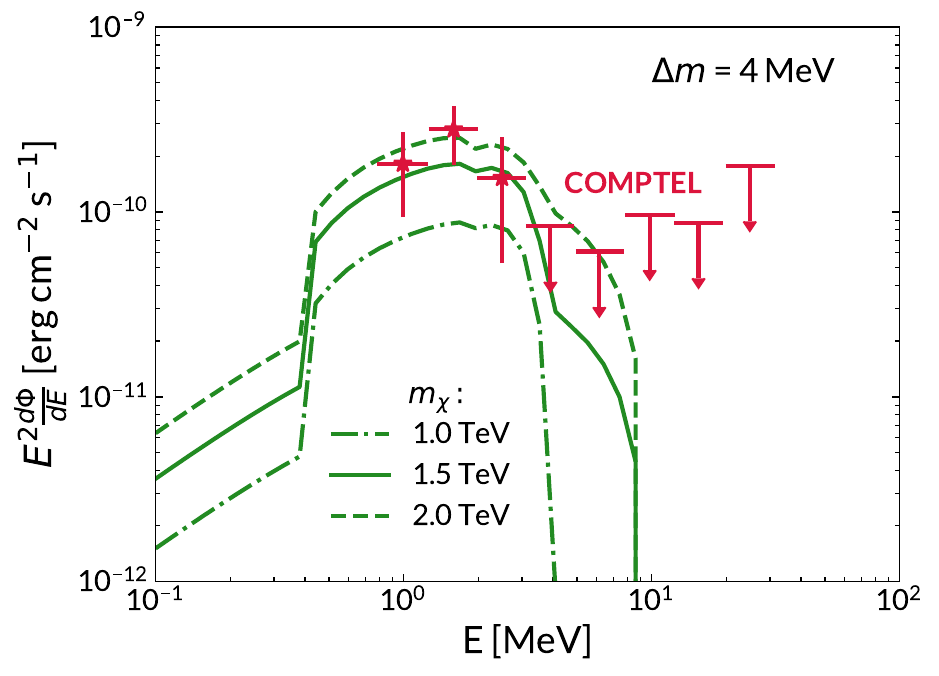}
    \includegraphics[width=0.49\linewidth]{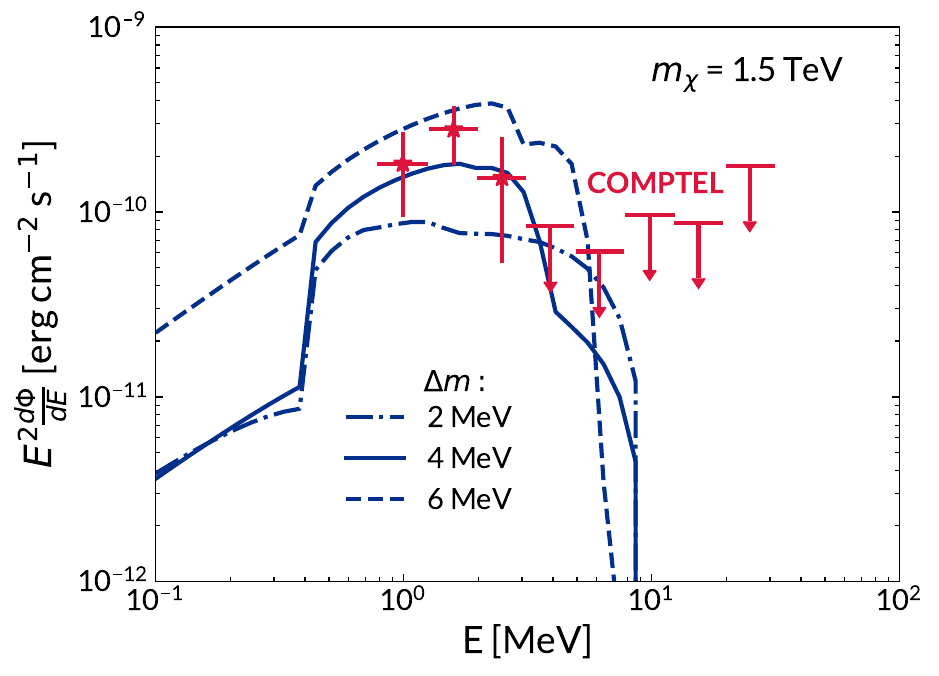}
    \includegraphics[width=0.49\linewidth]{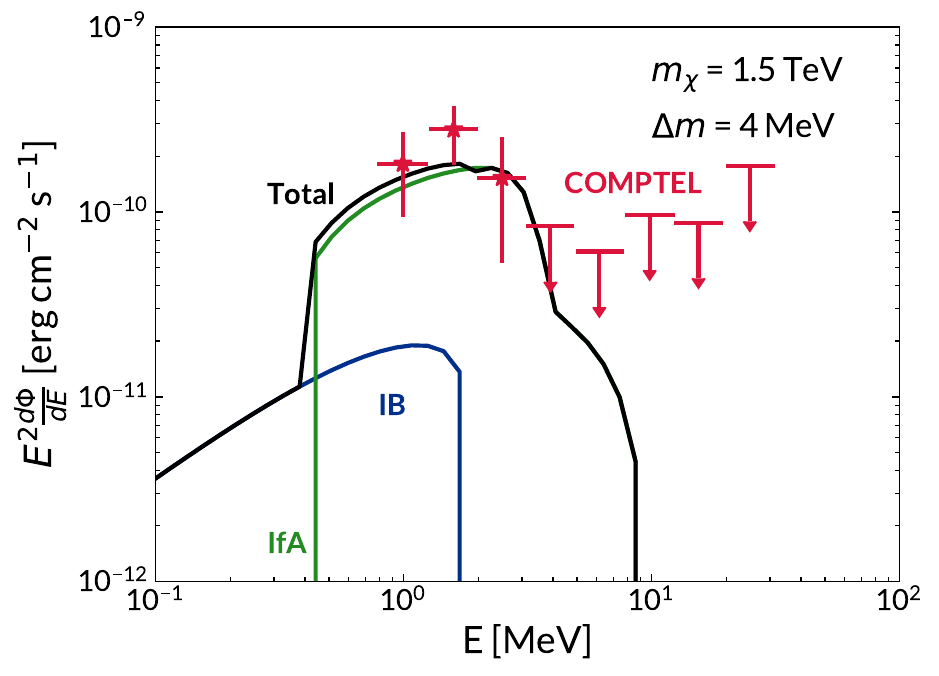}
    \caption{\textbf{Top-left:} COMPTEL IfA spectrum in the bulge region (red points) compared to XDM predictions for multiple $m_\chi$ with $\Delta m = 4$ MeV and $\sigma_{\textrm{mr}}$ fixed by the 511 keV fit. \textbf{Top-right:} Same, for fixed $m_\chi = 1.5$ TeV and varying $\Delta m$. \textbf{Bottom:} Decomposition of the XDM signal for $m_\chi = 1.5$ TeV and $\Delta m = 4$ MeV into internal bremsstrahlung (IB, blue) and in-flight annihilation (IfA, green). The sum (black) is compared to COMPTEL data.}
    \label{fig:IfA_best_fit}
\end{figure*}

\subsection{Propagation Uncertainties}

Fig.~\ref{fig:511Uncs} shows the impact of CR propagation uncertainties on the 511 keV morphology. We consider the ``Fast'', ``Default'' and ``Slow'' models of \citet{DelaTorreLuque:2023cef}, which bracket optimistic and pessimistic choices for the diffusion and energy-loss parameters. The resulting differences in the longitude and latitude profiles are modest, reflecting the low kinetic energy of the injected positrons: their transport is dominated by energy losses rather than spatial diffusion. The main conclusions regarding the XDM fit to SPI are therefore robust against reasonable variations of the propagation setup.

\subsection{In-Flight Annihilation Spectrum}

Using the best-fit cross sections $\sigma_{\mathrm{mr}}$ obtained from the 511 keV morphology for the models in Fig.~\ref{fig:511_best_fit_lat}, we compute the corresponding IfA spectra in the COMPTEL energy range. The top panels of Fig.~\ref{fig:IfA_best_fit} compare the all-sky XDM-induced emission to the bulge excess extracted by \citet{Knodlseder:2025pnx}. Internal bremsstrahlung (IB) associated with the outgoing $e^\pm$ is included following Eq.~(2) of \citet{Beacom:2004pe}, with $m_\chi$ replaced by $\Delta m/2$, the positron energy. The separate IB and IfA components for our fiducial model are shown in the bottom panel.

For our default benchmark $m_\chi = 1.5$ TeV and $\Delta m = 4$ MeV, the predicted XDM signal is remarkably consistent with the COMPTEL data within $1\sigma$ across the MeV bump. The integrated 511 keV photon flux in this case is
\begin{equation}
\Phi_{511} \simeq 1.1\times 10^{-4}\ \mathrm{ph\,s^{-1}\,cm^{-2}},
\end{equation}
while the best-fit value reported by \citet{Knodlseder:2025pnx} for a conventional 2.1 MeV annihilating DM scenario (including IB) is $\Phi_{511} = 1.4\times 10^{-4}\ \mathrm{ph\,s^{-1}\,cm^{-2}}$. Given the systematic uncertainties inherent to the 511 keV measurement and the modelling of the IfA signal, this level of agreement indicates that the XDM framework passes a non-trivial consistency check: the values of $\sigma_{\mathrm{mr}}$ required to fit the SPI morphology also predict an MeV continuum compatible with the COMPTEL excess.

The left panel of Fig.~\ref{fig:cNFWIfA} shows the IfA spectrum for the cNFW profile, for $m_\chi = 1.5$ and $2$ TeV. In this case the total IfA flux is lower than in the NFW case because the more concentrated DM profile forces the 511 keV fit to be obtained with a smaller cross section $\sigma_{\mathrm{mr}}$ in order not to overshoot the SPI flux in the inner region. This directly translates into a reduced IfA emission.

\subsection{Ionization of the Central Molecular Zone}

\begin{figure*}
    \centering
    \includegraphics[width=0.49\linewidth]{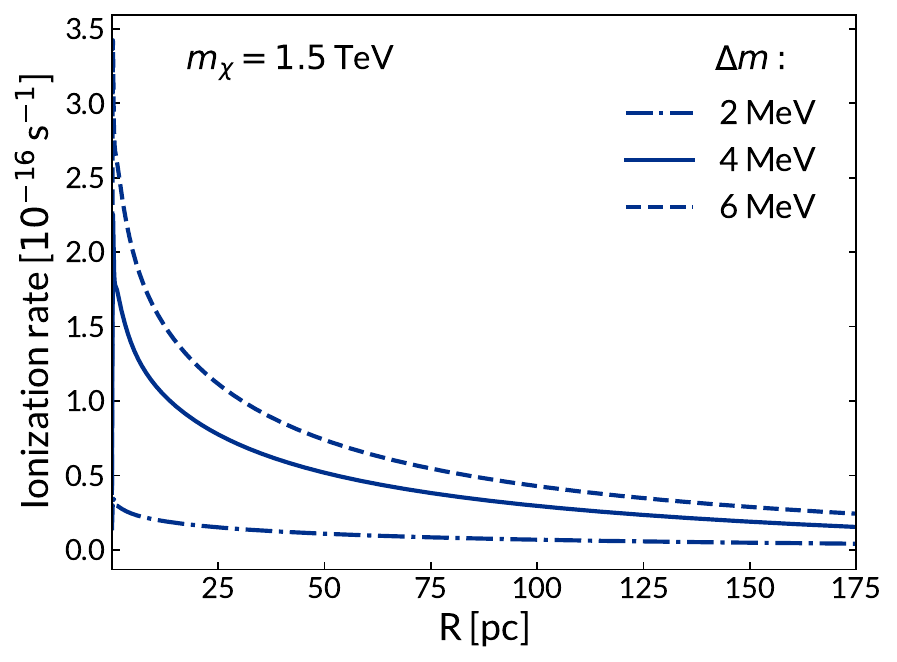}
    \includegraphics[width=0.49\linewidth]{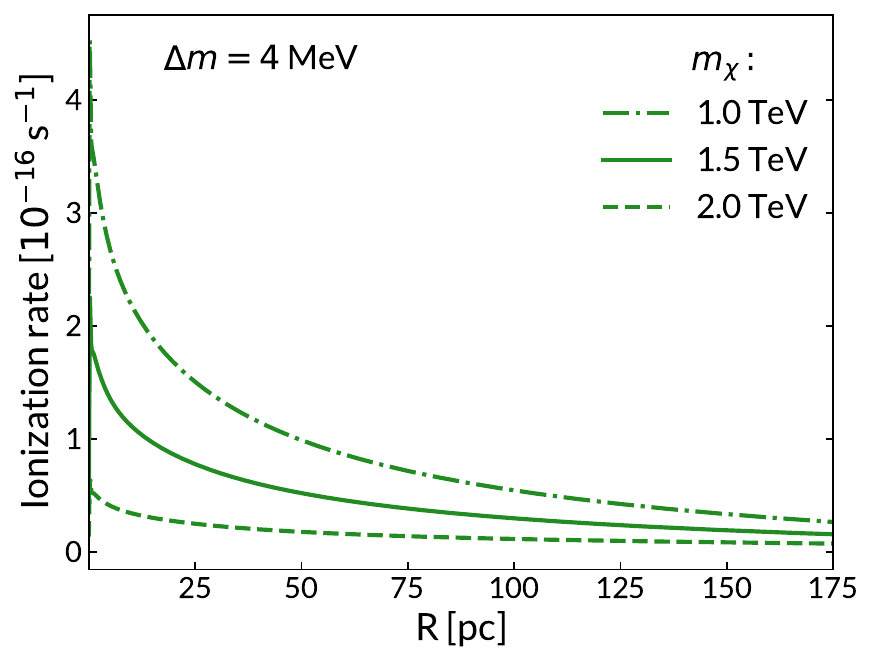}
    \caption{Radial profile of the H$_2$ ionization rate in the CMZ for XDM models as a function of cylindrical distance $R$ from the GC. Left: fixed $m_\chi = 1.5$ TeV and varying $\Delta m$. Right: fixed $\Delta m = 4$ MeV and varying $m_\chi$. All cases use NFW and the best-fit $\sigma_{\textrm{mr}}$ from the 511 keV morphology.}
    \label{fig:IonRate}
\end{figure*}

Fig.~\ref{fig:IonRate} shows the ionization rate $\zeta(R)$ in the CMZ obtained from Eq.~\eqref{eq:main} using the propagated XDM-induced $e^\pm$ fluxes from \textsc{DRAGON2}. For a fixed splitting $\Delta m = 4$ MeV and an NFW profile, the predicted ionization rates reach a few $\times 10^{-16}$ s$^{-1}$ inside the CMZ, above standard CR expectations and remaining relatively flat with radius, broadly consistent with the inferred profile \citep{Oka_2005,Oka_2020,Ravikularaman_2025} but still below the observed $\zeta_{\mathrm{H}_2}\gtrsim 10^{-15}\,\mathrm{s^{-1}}$. 

The normalization, however, does not fully saturate the observed CMZ ionization, which requires $\zeta_{\textrm{H}_2} \gtrsim 10^{-15}$ s$^{-1}$. Lower DM masses yield higher ionization rates, as they produce cuspier 511 keV and positron distributions. Similarly, cuspier DM profiles enhance the ionization near the GC, as shown in the right panel of Fig.~\ref{fig:cNFWIfA} for the cNFW case, but at the price of a less uniform radial profile and tighter constraints from the 511 keV morphology.

\citet{Obolentseva_2024} argued that ionization rates inferred for many molecular clouds across the Galaxy were biased high due to overestimates of their line-of-sight sizes and densities. This bias cannot by itself account for the extreme ionization levels in the CMZ, whose geometry and gas properties are better constrained, but it may lower the inferred CMZ ionization somewhat. Allowing for such a downward revision, the XDM-induced contribution could lie closer to the true ionization level and plausibly account for a non-negligible fraction of the excess. A more dedicated treatment of CR and DM transport in the complex environment of the CMZ is needed to sharpen this conclusion, but our results already show that the same XDM parameter space that fits the SPI and COMPTEL data naturally produces a significant, radially flat ionization component in the CMZ.

\section{Discussion and Conclusions}
\label{sec:conclusion}

We have shown that a minimal XDM framework can account for three correlated Galactic anomalies with a single mechanism. Inelastic upscattering $\chi\chi\!\to\!\chi\chi^*$ followed by $\chi^*\!\to\!\chi e^+e^-$ injects few-MeV positrons that (i) reproduce the 511~keV morphology seen by INTEGRAL/SPI, (ii) generate the 2–3~MeV IfA excess reported in reanalyzed COMPTEL data, and (iii) provide a substantial, radially flat contribution to the anomalous ionization in the CMZ. In this sense, MeV-band observations, ionization measurements and the Galactic positron budget are all linked by the same dark-sector process.

Using \textsc{DRAGON2}, we performed the first full CR propagation calculation of positrons in the XDM context. For $\Delta m \simeq 3$–$6$ MeV and TeV-scale $m_\chi$, the resulting 511~keV longitude and latitude profiles are in strong agreement with SPI data, while the corresponding IfA spectrum matches the COMPTEL MeV bump within current statistical and systematic uncertainties. The same parameter space yields CMZ ionization rates $\zeta \sim 10^{-16}$–$10^{-15}\,\mathrm{s^{-1}}$, larger than expected from standard CRs and with an approximately flat radial profile, although still below the nominal level inferred from molecular tracers. These late-time excitation rates can arise naturally from Sommerfeld enhancement in a light-mediator model and remain compatible with unitarity, thermal freeze-out and BBN constraints along the lines discussed in \citet{Finkbeiner:2007kk}.

A light boson of mass $\sim 10$–20 MeV coupling to $e^\pm$ can play the role of the mediator in such XDM scenarios. Intriguingly, this mass range overlaps with the mild ($\sim 2\sigma$) excess near $16.9$ MeV reported by the PADME experiment \cite{PADME:2025dla}, suggesting that forthcoming fixed-target searches may begin to probe exactly the mediator parameter space relevant for such Galactic MeV signatures. In parallel, future ``MeV gap'' missions such as \textit{COSI} \cite{Karwin_2023,Siegert_2020} will provide sharper line and continuum measurements and improved morphology, offering a direct test of the correlated line and continuum predictions of XDM.

Beyond the Milky Way, massive galaxy clusters provide another promising arena. Their high velocity dispersions make inelastic upscattering efficient throughout the halo, potentially leading to non-thermal X-ray or $\gamma$-ray emission and small contributions to intracluster-medium heating. Current limits on diffuse cluster emission already require any such heating to be subdominant to active galactic nucleus feedback, but the possibility that DM–DM scattering produces modest intracluster heating or mild SZ distortions remains open and may be probed by upcoming high-sensitivity X-ray and SZ surveys \cite{SimonsObservatory:2018koc,CMB-S4:2016ple}.

On cosmological scales, redshifted emission from XDM-induced $e^\pm$ injection and associated CMB energy-injection constraints could test scenarios with substantial early-universe heating. Velocity-dependent self-interactions of the type considered here also remain relevant for small-scale structure, and related dissipative dark-sector dynamics have been proposed as a channel for rapid black-hole seed formation at high redshift. These ideas can be confronted with next-generation observations of high-$z$ quasars and gravitational waves.

In summary, while extragalactic and cosmological signatures of XDM are still uncertain, the Galactic MeV window already unifies line emission, continuum excesses and molecular ionization in a coherent picture. If this interpretation is correct, the same few-MeV positrons that illuminate the GC in 511 keV photons are also leaving their imprint in the MeV continuum and in the chemistry of the CMZ, making MeV astrophysics a uniquely powerful probe of inelastic dark sectors.

\section{Acknowledgements}
SB is supported by the STFC under grant ST/X000753/1. DC acknowledges support from a Science and Technology Facilities Council (STFC) Doctoral Training Grant. 
PDL is supported by the Juan de la Cierva JDC2022-048916-I grant, funded by MCIU/AEI/10.13039/501100011033 European Union "NextGenerationEU"/PRTR. The work of PDL is also supported by the grants PID2021-125331NB-I00 and CEX2020-001007-S, both funded by MCIN/AEI/10.13039/501100011033 and by ``ERDF A way of making Europe''. PDL also acknowledges the MultiDark Network, ref. RED2022-134411-T. This project used computing resources from the National Academic Infrastructure for Supercomputing in Sweden (NAISS) under project NAISS 2024/5-666. This project used computing resources from King's College London CREATE system \cite{KCLCREATE2025}. For the purpose of open access, the authors have applied a Creative Commons Attribution (CC BY) license
to any Author Accepted Manuscript version arising from this submission.

\clearpage
\appendix
\section{Expected signals from a contracted NFW profile}
\label{app:cNFW}

\begin{figure*}[h]
    \centering  
    \includegraphics[width=0.49\linewidth]{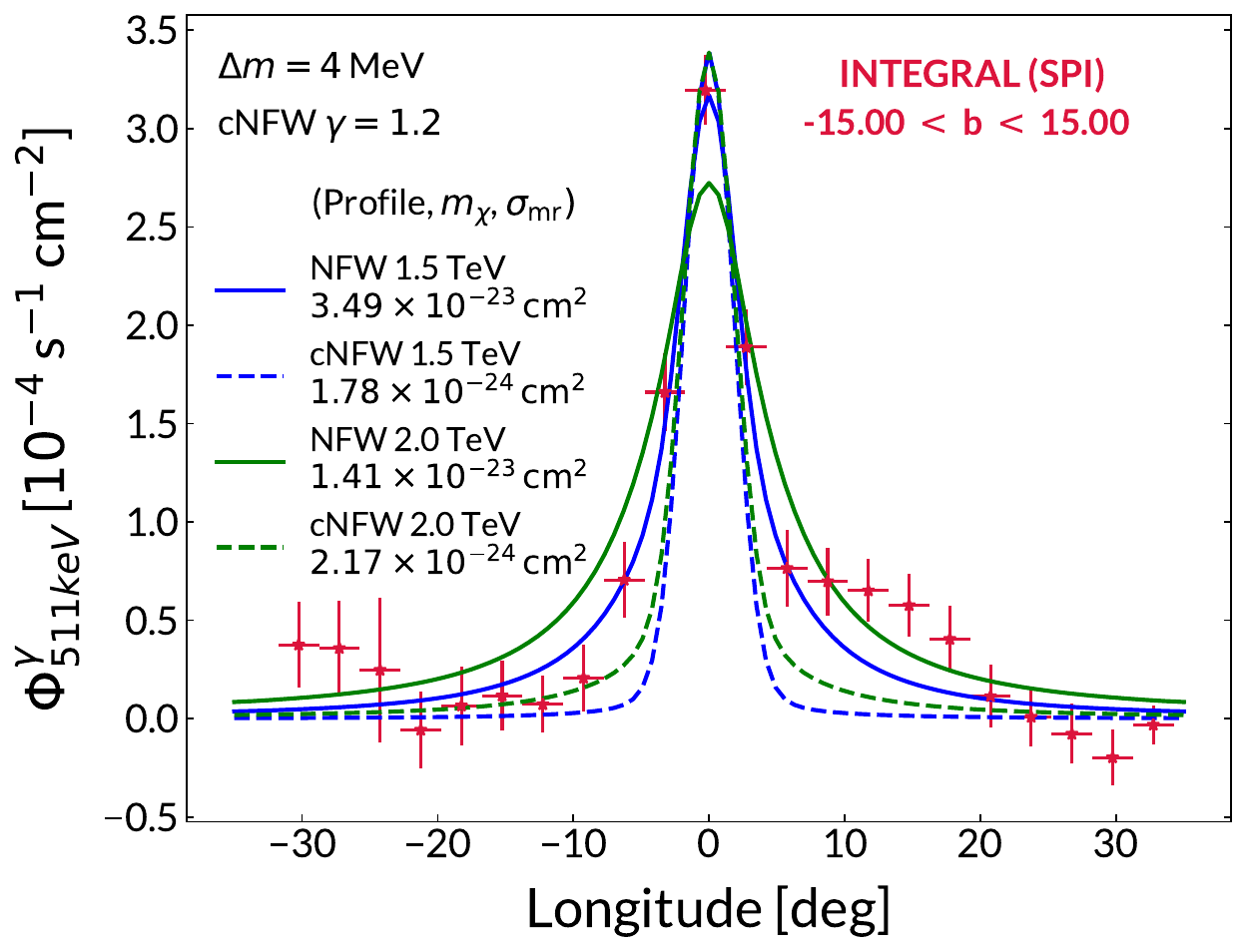}
    \includegraphics[width=0.49\linewidth]{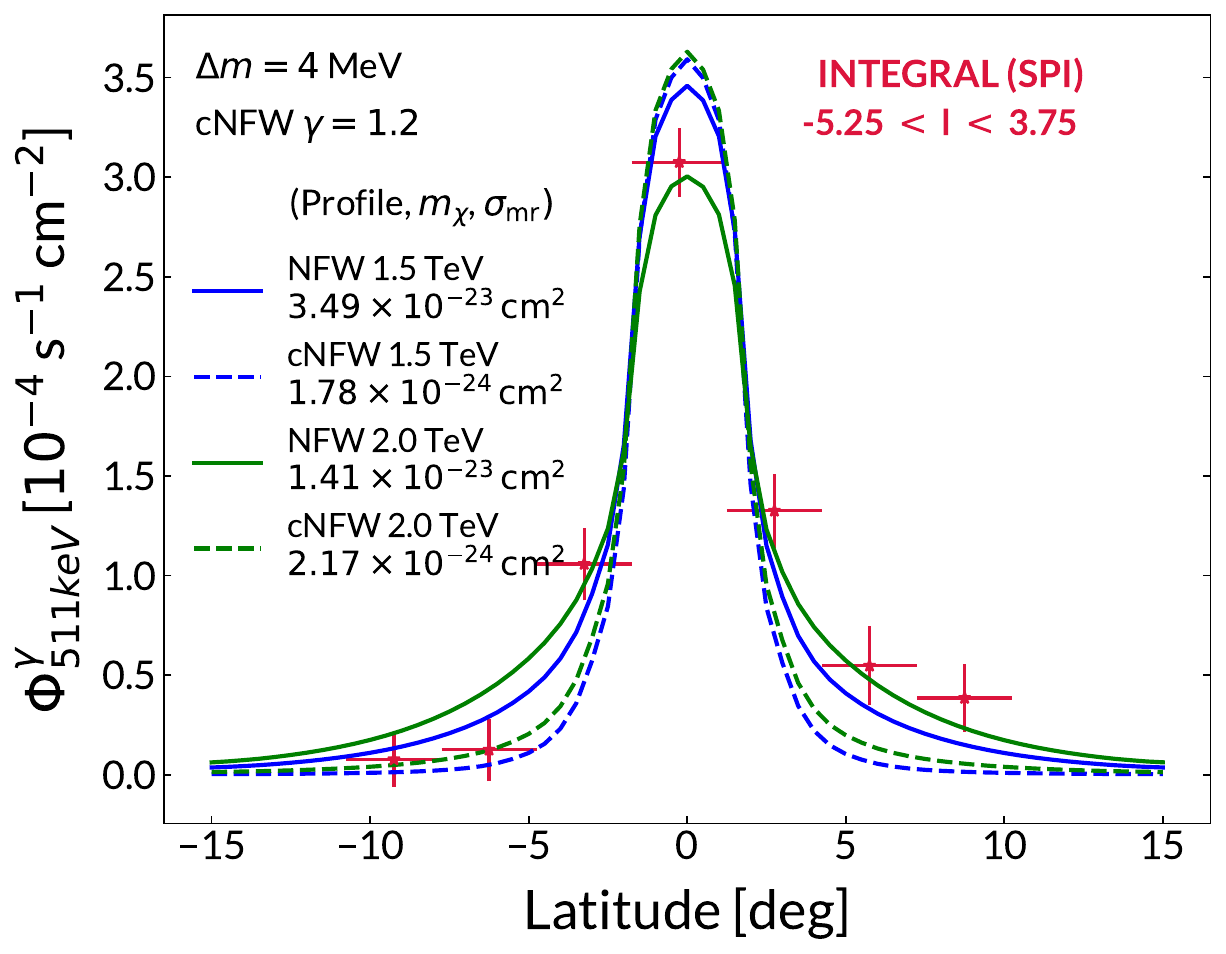}
    \caption{511 keV longitude (left) and latitude (right) profiles for XDM models with $m_\chi = 1.5$ TeV (blue) and $m_\chi = 2$ TeV (green), and fixed mass splitting $\Delta m = 4$ MeV, compared to INTEGRAL/SPI data (red points). Solid lines correspond to an NFW profile, dashed lines to a cNFW profile. The best-fit values of $\sigma_{\textrm{mr}}$ for each case are indicated in the legend.}
    \label{fig:511cNFW}
\end{figure*}

\begin{figure}[b]
    \centering
    \includegraphics[width=0.513\linewidth]{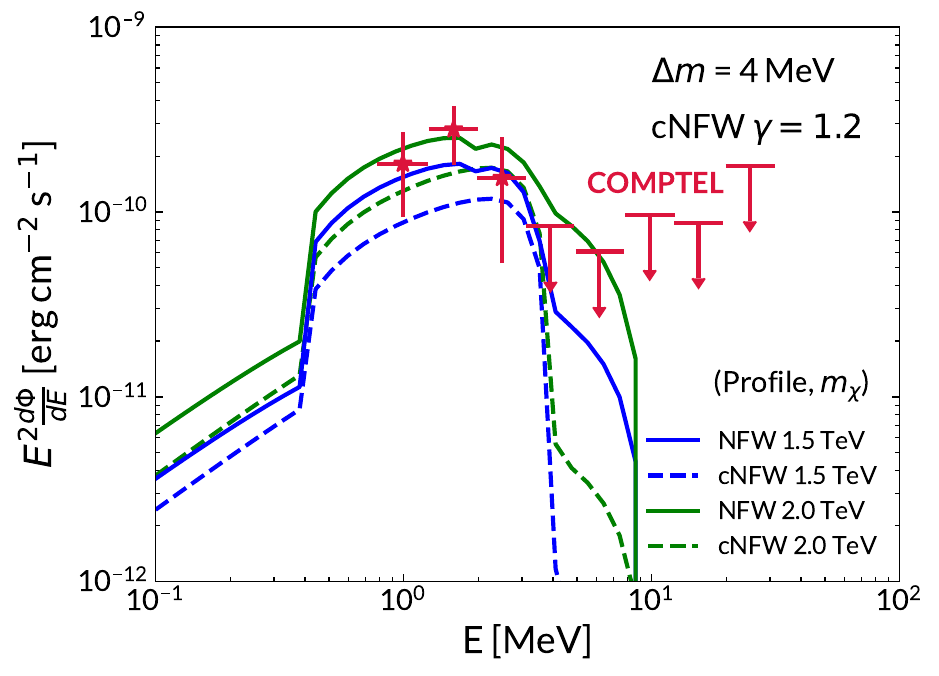}
    \caption{IfA flux measurements from COMPTEL (red points) compared to the XDM predictions for the same NFW and cNFW scenarios as in Fig.~\ref{fig:511cNFW}. The normalization in each case is set by the corresponding best-fit cross section $\sigma_{\textrm{mr}}$ obtained from the 511 keV morphology.}
    \label{fig:cNFWIfA}
\end{figure}

\begin{figure}
    \centering
    \includegraphics[width=0.477\linewidth]{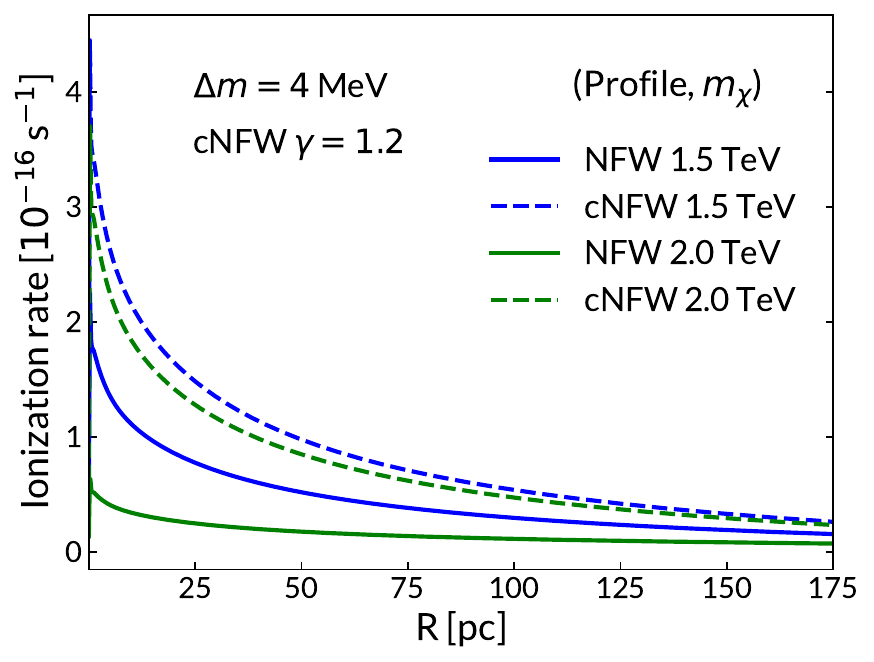}
    \caption{Radial profile of the H$_2$ ionization rate in the CMZ for the NFW and cNFW scenarios of Fig.~\ref{fig:511cNFW}. The normalization is again fixed by the best-fit $\sigma_{\textrm{mr}}$ from the 511 keV fit.}
    \label{fig:cNFWCMZ}
\end{figure}

In this appendix we quantify the impact of adopting a cNFW profile, with inner slope $\gamma = 1.2$, instead of the standard NFW halo. For each profile we refit the 511 keV morphology to obtain the corresponding best-fit cross section $\sigma_{\textrm{mr}}$, which is quoted in the legend of Fig.~\ref{fig:511cNFW} and used consistently in Figs.~\ref{fig:cNFWIfA} and \ref{fig:cNFWCMZ}.

\clearpage
\newpage
\bibliographystyle{apsrev4-1}
\bibliography{biblio.bib}

\end{document}